\documentclass[a4paper, twocolumn, 11pt, twoside]{article}

\usepackage[portuguese]{babel}
\usepackage{linguamatica-v14}

\submitted{\OCT{} 2017}
\accepted{\DEC{} 2017}

\title{Securing Bluetooth Low Energy: A Literature Review}

\titleEN{}

\author{
  Zhe Wang  
  \instituto{Harbin University of Science and Technology}
}

\begin{document}
\maketitle

%% Add the abstract in English
\begin{abstract}
  Bluetooth Low Energy (BLE) technology, operating within the widely used 2.4 GHz ISM band, stands as a cornerstone in modern wireless communication frameworks alongside its classic Bluetooth counterpart. This paper delves into the foundational aspects of BLE, excluding niche components, to explore its core functionalities and pivotal role in diverse connectivity needs. 
BLE's specialization in catering to low-power devices ensures optimal energy utilization, making it indispensable in IoT applications where energy efficiency is paramount. Its versatility finds applications across consumer electronics, industrial automation, and healthcare, ensuring reliability and efficiency in safety-critical systems and enhancing user convenience through remote control capabilities. 
However, the wireless nature of BLE interfaces exposes them to cybersecurity threats, necessitating robust security measures for mitigating risks such as sniffing, DoS attacks, and message injection. Continuous research and development efforts are essential to stay ahead of emerging threats and safeguard BLE-enabled systems and data.
\end{abstract}

%% Add the keywords in English

\section{Introduction}

Bluetooth technology operates within the widely used 2.4 GHz Industrial, Scientific, and Medical (ISM) band, encompassing two significant protocols: Bluetooth Basic Rate/Enhanced Data Rate (BR/EDR), commonly known as classic Bluetooth, and Bluetooth Low Energy (BLE), introduced in version 4.0. Despite their shared foundation, these protocols operate independently, serving diverse connectivity needs within the wireless ecosystem.

Within this expansive technological landscape, our discussion predominantly delves into the foundational aspects of BLE, excluding niche components such as Bluetooth Mesh, LE isochronous channels, and LE audio. This focus enables a comprehensive exploration of BLE's core functionalities and its pivotal role in modern wireless communication frameworks.

BLE stands out for its tailored approach towards low-power devices, catering to gadgets with constrained power supplies and limited computational resources. This specialization ensures optimal energy utilization, crucial for devices requiring prolonged battery life without sacrificing performance. The emphasis on efficiency makes BLE an indispensable component in the realm of Internet of Things (IoT), where power consumption is a critical consideration in device design and deployment.

The versatility of BLE extends across diverse sectors, finding applications in domains ranging from consumer electronics to industrial automation and healthcare. In safety-critical systems such as electronic locks, alarm systems, and medical devices, BLE's reliability and energy-efficient communication protocols play a pivotal role in ensuring seamless operation and user safety. Moreover, its integration with smartphones and laptops empowers users with remote control and monitoring capabilities, enhancing convenience and accessibility across various scenarios.

Despite its undeniable advantages, the wireless nature of BLE interfaces exposes them to potential cybersecurity threats. Attack vectors such as sniffing, denial-of-service (DoS), spoofing, message injection, and connection hijacking pose significant challenges to the security of BLE-enabled devices and networks. Mitigating these risks requires robust security measures, including device pairing schemes, encryption, authentication mechanisms, and address randomization. However, the effectiveness of these safeguards hinges on their proper implementation and adherence to best practices, highlighting the ongoing importance of cybersecurity awareness and diligence in the design and deployment of BLE solutions.

Furthermore, the evolving nature of technology necessitates a proactive approach towards addressing security vulnerabilities and enhancing resilience against emerging threats. Continuous research and development efforts are essential to stay ahead of malicious actors and safeguard the integrity and confidentiality of BLE-enabled systems and data.

In addition to security considerations, the longevity of BLE modules and devices in the market underscores the importance of backward compatibility with earlier protocol versions. This ensures seamless interoperability and support for legacy devices, facilitating smooth transitions and reducing potential disruptions in the evolving landscape of wireless connectivity.

\section{Background}

\subsection{Bluetooth Stack}
The Bluetooth stack, a fundamental architectural framework illustrated in Fig. 1, exhibits a hierarchical structure comprising two primary components: the controller and the host. This segmentation delineates the responsibilities and functionalities between lower-level and higher-level operations within the Bluetooth Low Energy (BLE) protocol.

\paragraph{Controller and Host Segmentation}:
The controller segment embodies the lower layers of the Bluetooth stack, which are primarily tasked with executing time-sensitive operations essential for the protocol's robust functioning. In contrast, the host encompasses layers responsible for orchestrating more complex and higher-level tasks, leveraging the outputs from the controller to facilitate seamless communication and interaction within the BLE ecosystem. This division serves to streamline the management of critical operations while providing the necessary abstraction for the implementation of advanced features and functionalities.

\paragraph{Host Controller Interface (HCI)}:
A pivotal component facilitating communication between the controller and the host is the host controller interface (HCI). Through this interface, the host issues commands to the controller, directing its activities, while the controller transmits data and pertinent information to the upper layers of the host. This bidirectional communication pathway enables efficient coordination and synchronization between the controller and the host, ensuring optimal performance and interoperability across the BLE protocol stack.

\paragraph{Application Interaction}:
The application layer, situated atop the host segment of the BLE stack, serves as the primary interface for user interaction and application development. By interfacing with the upper layers of the host, applications can leverage the functionalities and services provided by the BLE protocol stack to implement a wide array of wireless communication and networking applications. This seamless integration empowers developers to harness the capabilities of BLE for diverse use cases, ranging from IoT applications to smart devices and beyond.

\begin{figure}
    \centering
    \includegraphics[width=\linewidth]{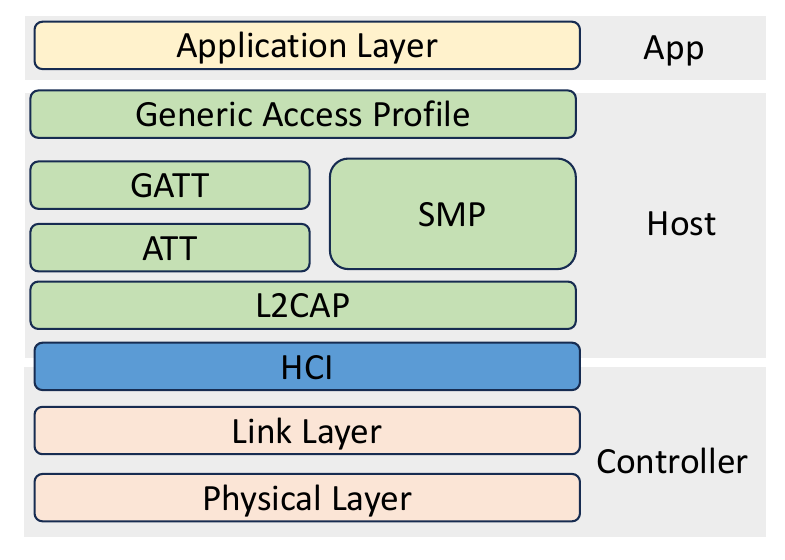}
    \caption{Bluetooth Stack}
    \label{fig:enter-label}
\end{figure}

\subsection{Exploration of Protocol Layers}
 
Within the BLE protocol stack, various layers play distinct yet interconnected roles in facilitating efficient communication and data exchange. The subsequent sections delve deeper into the functionalities and intricacies of these layers, shedding light on their contributions to the overall operation and performance of the Bluetooth technology.

\begin{itemize}
    \item \textbf{Physical Layer}:
At the foundational level of the BLE stack lies the physical layer (PHY), which defines the specific capabilities and characteristics of the BLE radio and analog circuitry. Operating within the internationally allocated ISM band spanning from 2.402 GHz to 2.480 GHz, the PHY establishes the groundwork for wireless communication by partitioning the available spectrum into a series of radio frequency (RF) channels. This segmentation enables concurrent transmission and reception of data, ensuring efficient utilization of the available bandwidth while mitigating interference from other wireless technologies operating within the same frequency range.

\item \textbf{Link Layer}:
Sitting atop the physical layer, the link layer (LL) assumes responsibility for managing low-level operations essential for establishing and maintaining communication links between BLE devices. From reception and processing of protocol data units (PDUs) to the execution of security-related functions such as encryption and decryption, the LL plays a pivotal role in ensuring the integrity, security, and reliability of data exchange within the BLE network. Additionally, the LL maintains vital device identifiers, such as the Bluetooth device address (BD\_ADDR), facilitating device discovery and differentiation within the network topology.

\end{itemize}

By exploring the intricate layers of the Bluetooth stack in greater detail, we gain a deeper understanding of the underlying mechanisms driving the seamless and efficient operation of Bluetooth Low Energy technology, paving the way for innovative applications and advancements in wireless communication.

\subsection{Security Manager}

The security manager (SM) layer plays a crucial role in overseeing the intricate facets of Bluetooth security, encompassing a spectrum of tasks vital for safeguarding data transmission. This pivotal layer can be delineated into two primary components: a robust cryptographic toolbox and a suite of methodologies dedicated to pairing and key exchange protocols.

Within the cryptographic toolbox lie a myriad of essential functions, facilitating hash calculations and the generation of cryptographic keys integral to the secure operation of Bluetooth-enabled devices. These cryptographic functions serve as the backbone of Bluetooth security, ensuring the confidentiality, integrity, and authenticity of data exchanged over Bluetooth connections.

Moreover, the SM layer orchestrates sophisticated pairing mechanisms and key exchange procedures, vital for establishing secure communication channels between devices. These methods are designed to authenticate the identity of participating devices, mitigate potential security threats, and facilitate the seamless exchange of encryption keys to fortify the confidentiality of transmitted data.

In essence, the multifaceted responsibilities entrusted to the security manager (SM) layer underscore its indispensable role in fortifying the security posture of Bluetooth-enabled ecosystems, offering robust protection against unauthorized access, data breaches, and other malicious exploits.

\section{Bluetooth Security and Privacy}

\subsection{Bluetooth Security}

Bluetooth pairing is a fundamental process in establishing secure communication between two or more Bluetooth-enabled devices. It involves the mutual authentication of devices and the exchange of encryption keys to secure the data transmitted between them. The pairing process typically occurs when two devices are attempting to establish a connection for the first time or when they haven't been previously paired. 
The Bluetooth pairing process typically involves the following steps:

\begin{itemize}
    \item 
\textbf{Initiation}: One device initiates the pairing process by broadcasting its availability and readiness to establish a connection. This device is often referred to as the "initiator" or "master."
\item 
\textbf{Discovery}: The initiating device scans for nearby Bluetooth devices within its range. It identifies potential devices with which it can establish a connection.
\item 
\textbf{Device Selection}: The initiating device selects the specific device with which it wants to pair from the list of discovered devices.
\item 
\textbf{Authentication}: Once the devices have been identified, they undergo an authentication process to ensure that they are legitimate and authorized to communicate with each other. Authentication mechanisms may include passkeys, PINs, or other security codes entered by the users of the devices.
\item 
\textbf{Pairing Request}: The initiating device sends a pairing request to the selected device, indicating its intent to establish a secure connection.
\item 
\textbf{Acceptance}: The selected device receives the pairing request and prompts the user to confirm the pairing. If the user accepts, the devices proceed to establish a secure connection.
\item 
\textbf{Key Exchange}: During the pairing process, the devices exchange encryption keys that will be used to encrypt and decrypt data transmitted between them. This ensures that the data remains confidential and secure during transmission.
\item 
\textbf{Connection Establishment}: Once the pairing process is successfully completed, the devices establish a connection, enabling them to exchange data securely.
\end{itemize}

It's worth noting that Bluetooth pairing can occur using different pairing methods, such as Numeric Comparison, Passkey Entry, or Just Works. The method used depends on the security requirements of the devices involved and the preferences of the users.

\subsection{Bluetooth Privacy}

Bluetooth Identity Resolution is a feature introduced in Bluetooth version 4.2 that enhances privacy and security in Bluetooth Low Energy (BLE) connections. It is particularly important in scenarios where multiple devices interact within a shared environment, such as in the context of the Internet of Things (IoT) or proximity-based services.

The primary objective of Bluetooth Identity Resolution is to prevent tracking or identification of devices based solely on their Bluetooth MAC addresses. Traditionally, Bluetooth devices broadcast their MAC addresses, which can be captured and used for tracking purposes, potentially compromising user privacy and security.

Bluetooth Identity Resolution addresses this issue by introducing the concept of Resolvable Private Addresses (RPAs) and the associated resolving process. Here's how it works:

\paragraph{Resolvable Private Addresses (RPAs)}: Devices operating in BLE can generate random, periodically changing MAC addresses known as Resolvable Private Addresses. These addresses are designed to be untraceable, as they change frequently and cannot be linked to a specific device over time. RPAs are used for advertising and connection establishment, enhancing privacy by preventing unauthorized tracking based on MAC addresses.

\paragraph{Identity Resolving Key (IRK)}: Each BLE device maintains an Identity Resolving Key (IRK), which is a shared secret key used to generate and resolve RPAs. The IRK is stored securely within the device and is only shared with trusted parties, such as paired devices.

\paragraph{Resolving Process}: When a BLE device encounters another device broadcasting an RPA, it can attempt to resolve the RPA back to the corresponding static identity using the IRK. If the device has the necessary IRK, it can perform the resolution process and identify the other device without compromising privacy. This allows devices to maintain secure connections with known peers while preserving privacy in public or shared environments.

By implementing Bluetooth Identity Resolution, BLE devices can enjoy improved privacy protection without sacrificing connectivity or functionality. It enables secure communication while mitigating the risk of tracking or unauthorized surveillance based on MAC addresses, thus fostering trust and confidence in Bluetooth-enabled applications and services.

\section{Bluetooth Attacks} 

\textbf{Passive Sniffing}: In passive sniffing attacks, the attacker intercepts Bluetooth communication passively, without actively participating in the connection. This involves monitoring Bluetooth traffic using specialized hardware or software tools capable of capturing Bluetooth packets exchanged between devices. Passive sniffing is often used to gather information about device activity, identify potential vulnerabilities, or extract sensitive data for further exploitation.

\paragraph{Active Sniffing}: Active sniffing attacks involve the attacker actively participating in Bluetooth communication by impersonating a legitimate device or manipulating connection parameters. By injecting malicious packets into the communication stream or forcing devices to re-establish connections, the attacker can gain access to sensitive information exchanged between devices. Active sniffing attacks are more intrusive and can potentially disrupt legitimate communication between devices.

\paragraph{Man-in-the-Middle (MITM) Attacks}: MITM attacks occur when the attacker intercepts and alters communication between two parties, acting as an intermediary without their knowledge. In the context of Bluetooth, MITM attacks can be used to intercept authentication credentials, session keys, or other sensitive data exchanged during the pairing process. By positioning themselves between the communicating devices, the attacker can capture and manipulate data packets, compromising the confidentiality and integrity of the communication.

\paragraph{Brute Force Attacks}: Brute force attacks involve systematically trying all possible combinations of authentication credentials, such as PINs or passkeys, to gain unauthorized access to a Bluetooth device. This type of attack is often used when attempting to pair with a device that has weak or default credentials. By repeatedly guessing authentication codes, the attacker can eventually gain access to the device and its data.

\paragraph{Bluejacking}: Bluejacking is a relatively harmless form of Bluetooth attack where an attacker sends unsolicited messages or files to nearby Bluetooth-enabled devices. The goal of bluejacking is typically to annoy or prank the recipient rather than to cause harm. However, it can still be used to disrupt normal device operation or to gather information about nearby devices.

\paragraph{Bluesnarfing}: Bluesnarfing is a more serious Bluetooth attack that involves unauthorized access to information stored on a Bluetooth-enabled device, such as contacts, emails, text messages, or multimedia files. Attackers exploit vulnerabilities in the Bluetooth protocol or implementation to gain access to the device's data without the user's knowledge or consent.

\paragraph{Bluebugging}: Bluebugging is an advanced Bluetooth attack that allows an attacker to take control of a Bluetooth-enabled device without the user's knowledge. By exploiting vulnerabilities in the Bluetooth protocol or device firmware, attackers can remotely execute commands, make calls, send messages, or access sensitive information stored on the device. Bluebugging attacks typically require specialized tools and technical expertise to execute.

\paragraph{Denial of Service (DoS) Attacks}: DoS attacks target the availability of Bluetooth-enabled devices or networks by flooding them with a high volume of malicious traffic or by exploiting vulnerabilities to crash or freeze the devices. DoS attacks can disrupt Bluetooth communication, rendering devices unable to connect or causing them to malfunction.

\paragraph{Impersonation Attacks}: In impersonation attacks, attackers attempt to impersonate legitimate Bluetooth devices or services to deceive users or other devices into connecting to them. By masquerading as a trusted device or service, attackers can intercept communication, steal sensitive information, or execute other malicious actions.

\paragraph{Bluetooth Malware and Exploits}: Malicious software or exploits specifically designed to target Bluetooth-enabled devices can compromise their security and integrity. Bluetooth malware may infect devices through malicious files or applications, exploiting vulnerabilities to gain unauthorized access, steal data, or perform other malicious activities.

\section{Bluetooth Defenses}

Defending against Bluetooth-related attacks requires a multi-layered approach that combines technical controls, security best practices, and user awareness. Here are some key defenses to mitigate the risks associated with Bluetooth attacks:

\paragraph{Authentication and Encryption}: Enable strong authentication and encryption mechanisms, such as Bluetooth Secure Simple Pairing (SSP) or Secure Connections. This ensures that only authorized devices can connect to each other and that data transmitted over Bluetooth connections remains confidential and secure.

\paragraph{Firmware and Software Updates}: Regularly update device firmware and software to patch known vulnerabilities and ensure that devices are protected against the latest security threats. This includes updating operating systems, Bluetooth drivers, and application software on all Bluetooth-enabled devices.

\paragraph{Disable Unused Bluetooth Features}: Disable unnecessary Bluetooth services or features that are not required for device functionality. This reduces the attack surface and minimizes the risk of exploitation by disabling features that could be used as attack vectors.

\paragraph{Use Strong PINs and Passkeys}: When pairing Bluetooth devices, use strong and unique Personal Identification Numbers (PINs) or passkeys to prevent brute force attacks. Avoid using default or easily guessable PINs, as they can be easily exploited by attackers.

\paragraph{Bluetooth Sniffing Detection}: Deploy Bluetooth sniffing detection tools to monitor for suspicious activity and unauthorized access attempts. These tools can help detect and alert administrators to potential Bluetooth attacks in real-time, allowing for prompt response and mitigation.

\paragraph{User Education and Awareness}: Educate users about the risks associated with Bluetooth-enabled devices and how to protect against potential threats. This includes teaching users how to recognize suspicious Bluetooth activity, avoid connecting to unknown or untrusted devices, and securely configure Bluetooth settings on their devices.

\paragraph{Bluetooth Range Limitation}: Limit the range of Bluetooth signals by reducing transmission power or using directional antennas. This helps mitigate the risk of eavesdropping and unauthorized access by limiting the distance over which Bluetooth signals can be intercepted.

\paragraph{Implement Network Segmentation}: Segment Bluetooth-enabled devices into separate network zones or VLANs to isolate them from critical systems and sensitive data. This helps contain potential Bluetooth attacks and prevents attackers from gaining unauthorized access to sensitive resources.

\paragraph{Bluetooth Device Management}: Implement centralized Bluetooth device management solutions to monitor and control Bluetooth-enabled devices within the organization. This includes inventory management, configuration management, and remote device management capabilities to ensure that Bluetooth devices are properly configured and securely maintained.

\vspace{2mm}

By implementing these defenses, organizations can strengthen their Bluetooth security posture and reduce the risk of Bluetooth-related attacks. However, it's important to recognize that no defense is foolproof, and a layered approach that combines multiple security measures is essential to effectively mitigate the risks associated with Bluetooth technology.

\section{Related Work}

\subsection{Bluetooth Security}

In the realm of Bluetooth security~\cite{zhang2020bless,zhang2019security,zhang2020breaking}, a rich tapestry of research endeavors has illuminated various facets of vulnerabilities and potential exploits, shedding light on the intricate landscape of wireless communication protocols. Previous investigations spearheaded by luminaries like Mike Ryan \cite{ryan2013bluetooth} and Rosa \cite{rosa2013bypassing} have uncovered glaring security flaws in Passkey Entry for LE legacy connections, leveraging sophisticated tools such as crackle. While these seminal studies laid the groundwork for understanding vulnerabilities in Bluetooth pairing mechanisms, our research extends beyond the confines of legacy connections to encompass the more recent and fortified BLE versions 4.2 and 5.x.

Venturing into the contemporary terrain of Bluetooth Low Energy (BLE), a myriad of challenges and vulnerabilities have surfaced, prompting a concerted effort to unravel their intricacies and devise robust countermeasures. Noteworthy among these challenges is the brute-force cracking of BLE temporary keys, as demonstrated by Zegeye et al. \cite{zegeye2015exploiting}, presenting a formidable obstacle to the security of BLE-enabled devices. Furthermore, the work of Dazhi Sun et al. \cite{sivakumaran2018low} illuminates the perils of reusing passkeys in BLE contexts, a practice vehemently discouraged due to its susceptibility to exploitation.

Beyond the realm of BLE, authentication vulnerabilities in Bluetooth Classic have also come under scrutiny, courtesy of Antonioli et al. \cite{Antonioli2020BIAS}, unraveling the potential for downgrading the Secure Connections protocol and raising concerns about the integrity of legacy Bluetooth implementations. Concurrently, the research efforts of Jasek et al. \cite{jasek2016gattacking} pivot towards exploring attacks between Bluetooth smart devices and their companion mobile apps, accentuating the evolving threat landscape in the realm of Bluetooth security.

Expanding the scope of inquiry, several studies have delved into the intricacies of reverse engineering specific Bluetooth-enabled products, unveiling vulnerabilities lurking within their communication protocols. Cyr et al. \cite{cyr2014security} conducted a meticulous security analysis of wearable fitness devices, employing reverse engineering techniques to dissect the intricacies of BLE communication traffic and uncover vulnerabilities ripe for exploitation. Similarly, Zhang et al. \cite{zhang2017security} embarked on a journey to dissect the commands emanating from popular smart wristbands, unveiling vulnerabilities susceptible to replay and man-in-the-middle (MITM) attacks, thereby underscoring the imperative for robust security measures in the burgeoning landscape of wearable technology.

The specter of vulnerabilities extends beyond individual devices to encompass broader systemic flaws, which exposed fundamental flaws in BLE implementations, posing a pervasive threat to the security of interconnected devices. Additionally, studies by William et al. \cite{oliff2017evaluating} and Melamed et al. \cite{melamed2018active} delve into the realm of spoofing and MITM attacks between Bluetooth smart devices and their mobile apps, highlighting the need for heightened vigilance and robust security protocols.

Moreover, the discourse surrounding Bluetooth security has been enriched by the contributions of scholars such as Muhammad Naveed et al. \cite{NaveedZDWG14}, Xu et al. \cite{xu2019badbluetooth}, and Sivakumaran et al. \cite{sivakumaran2018attacks}, each offering unique insights into the multifaceted nature of Bluetooth vulnerabilities and potential mitigation strategies. Notably, the pioneering work of Zuo et al. \cite{zuo2019automatic} underscores the efficacy of fingerprinting via UUIDs in identifying vulnerable IoT devices employing insecure pairing mechanisms, paving the way for proactive security measures to safeguard against emerging threats in the ever-evolving landscape of Bluetooth technology.   Mohit et al.~\cite{jangid2023extrapolating} propose a detailed formal analysis of the Passkey Entry (PE) protocol in Bluetooth security, uncovering known and new attacks, and verifying fixes for vulnerabilities, ultimately demonstrating the confidentiality and authentication properties of PE through an inductive base model. 
Wang et al.~\cite{wang2023one} addressed security concerns stemming from diverse platform implementations.

\subsection{Bluetooth Privacy}

Over the span of the last two decades, the landscape of cybersecurity has witnessed a significant surge in the emergence of Bluetooth device tracking attack methodologies, each aimed at exploiting vulnerabilities inherent in the Bluetooth protocol. This proliferation of attack proposals, documented in a plethora of research endeavors, underscores the growing concern surrounding the security implications of Bluetooth technology in various contexts. Researchers have devoted considerable efforts to understanding and mitigating these threats, as evidenced by a diverse array of studies and proposals in the field.

The genesis of these tracking attacks can be traced back to the early 2000s, with seminal works such as BlueTrack \cite{haase2004bluetrack} and BLEB \cite{issoufaly2017bleb}, which laid the groundwork for subsequent research in this domain. These pioneering efforts primarily focused on exploiting vulnerabilities associated with Bluetooth devices utilizing public addresses, thereby facilitating the tracking of such devices through the interception of advertising packets. Building upon these foundational studies, researchers have since embarked on a quest to unravel the intricacies of Bluetooth device tracking in increasingly sophisticated scenarios.

One notable area of investigation pertains to the challenges posed by the proliferation of address randomization techniques employed by Bluetooth devices. Address randomization, aimed at enhancing user privacy and thwarting tracking attempts, presents a formidable obstacle for adversaries seeking to persistently track Bluetooth devices. The dynamic nature of randomized MAC addresses, coupled with the transient nature of device visibility in Bluetooth networks, complicates traditional tracking methodologies and necessitates novel approaches to address this evolving threat landscape.

In response to these challenges, researchers have devised innovative strategies to circumvent address randomization and enhance the efficacy of Bluetooth device tracking. For instance, Marco et al. \cite{cominelli2020even} delved into the realm of Bluetooth classic devices devoid of address randomization, leveraging leaked information from frame encoding to facilitate tracking. Similarly, Ludant et al. \cite{ludantlinking} proposed a novel approach to linking and tracking BLE addresses by collecting Bluetooth Classic (BT) addresses, thereby offering insights into the interplay between different Bluetooth protocols and their implications for tracking.

Furthermore, advancements in the field have led to the development of sophisticated tools and frameworks tailored to augmenting Bluetooth device tracking capabilities. Solutions such as BLE-Guardian \cite{fawaz2016protecting} and BLEScope \cite{zuo2019automatic} exemplify the ongoing efforts to combat the challenges posed by randomized MAC addresses, offering users the flexibility to customize packet accessibility and employing fingerprinting techniques to identify vulnerable Bluetooth devices.

Despite these strides in mitigating Bluetooth device tracking attacks, the threat landscape continues to evolve, with adversaries devising increasingly sophisticated tactics to exploit vulnerabilities in Bluetooth implementations. Targeted tracking attacks against specific device types, such as Apple devices \cite{celosia2020discontinued, martin2019handoff, stute2021disrupting, stute2019billion} and wearable fitness trackers \cite{das2016uncovering}, underscore the need for continuous vigilance and innovation in the realm of Bluetooth security. Zhang et al.~\cite{zhang2022good} identify vulnerabilities in BLE technology, where allowlisting intended for connection restriction inadvertently enables device tracking, compounded by flaws in the MAC address randomization scheme susceptible to replay attacks

In conclusion, the proliferation of Bluetooth device tracking attacks underscores the imperative for ongoing research and collaboration to address the evolving threats posed by Bluetooth technology. By leveraging innovative approaches and developing robust countermeasures, researchers can bolster the security posture of Bluetooth-enabled devices and safeguard against emerging threats in an increasingly interconnected world.

\subsection{IoT Security and Privacy}

Fuzzing has emerged as a prevalent and potent technique for uncovering vulnerabilities within software systems, offering a proactive approach to identifying weaknesses before they can be exploited by malicious actors. In recent years, the proliferation of Internet of Things (IoT) devices has introduced a new frontier for fuzzing-based vulnerability detection and defenses~\cite{luo2022security,liu2020manually,luo2020runtime,pearson2022fume,pearson2020sic,shao2021peripheral,pearson2019misconception,luo2018security}. A plethora of research endeavors has been undertaken to address the unique challenges posed by the diverse landscape of IoT devices, with several notable studies paving the way for advancements in this domain \cite{issta::2022::zheng, mobisys::2023::ma, usenix::2023::greenhouse, SP::2021::diane, NDSS::chen::2018, CCS::2021::snipuzz, firm-alf::2019::usenix, usenix::fuzzware::2022}.
These efforts can broadly be categorized into two distinct approaches: black box IoT device fuzzing and grey box IoT device fuzzing. Black box fuzzing methods, exemplified by studies such as those by Chen et al. \cite{NDSS::chen::2018} and others \cite{CCS::2021::snipuzz, SP::2021::diane, mobisys::2023::ma}, tackle the challenge of closed-source and architecturally diverse IoT devices by treating them as opaque entities. In this approach, fuzzing is conducted by dispatching messages to the target devices from a controller, with subsequent analysis of side-channel cues such as network behavior to detect potential vulnerabilities.
Chen et al. \cite{NDSS::chen::2018} present a groundbreaking black box IoT device fuzzing method, leveraging companion Apps to identify network-related or data-encoding method-related UI components. By intercepting and manipulating user input, valid fuzzing packets are generated for IoT devices. However, while these black-box testing methodologies have proven effective within local networks, they often fall short when confronted with remote control scenarios involving cloud servers, highlighting a significant area for further exploration and refinement \cite{firm-alf::2019::usenix, usenix::fuzzware::2022, usenix::2023::greenhouse}.
In contrast, grey box IoT device fuzzing approaches, as explored by researchers such as Zheng et al. \cite{issta::2022::zheng}, seek to strike a balance between the black box and white box paradigms by incorporating limited knowledge about the internal workings of the target IoT devices. This hybrid approach enables more targeted and nuanced fuzzing strategies, leveraging insights gleaned from partial analysis of device architecture and behavior to optimize vulnerability detection. Gao et al.~\cite{gao2019microcontroller} examine security concerns in Microcontroller (MCU) based IoT firmware, addressing the vulnerabilities in contemporary firmware upgrade models through case studies, including an attack validation on a popular air quality sensor from PurpleAir and investigation of a secure firmware upgrade system prototype on an ATmega1284P chip. 

\section{Conclusion}

In conclusion, Bluetooth Low Energy (BLE) technology has emerged as a foundational element in modern wireless communication ecosystems, offering tailored solutions for low-power devices while catering to diverse connectivity needs across various sectors. Its emphasis on energy efficiency makes it indispensable in the realm of IoT, where prolonged battery life is essential. Moreover, its versatility finds applications in safety-critical systems and enhances user convenience through remote control capabilities. 
However, the inherent wireless nature of BLE interfaces exposes them to cybersecurity threats, necessitating robust security measures to mitigate risks and safeguard systems and data. Continuous research and development efforts are crucial to stay ahead of emerging threats and ensure the integrity and confidentiality of BLE-enabled solutions.

\bibliography{references.bib}

\end{document}